\documentclass[a4paper, 12pt, openany, oneside]{article}
\usepackage[cp1251]{inputenc}
\usepackage[english]{babel}
\usepackage{ graphics,amsfonts, amsmath,bm,amssymb, array}
\usepackage[dvips]{graphicx}

\renewcommand{\@biblabel}[1]{#1.\hfill}

\newcommand{\intl}{\int\limits}

\renewcommand{\Re}{\mathop{\rm Re\,}}

\newcommand{\Res}{\mathop{\rm Res\,}}
\makeatother {

\begin{document}
\thispagestyle{empty} \large

\renewcommand{\refname}{\normalsize\begin{center}\bf
REFERENCES\end{center}}

\begin{center}
{\bf Interaction of the Electromagnetic p-Wave with Thin Metal Film
in the Field of Resonant Frequencies}
\end{center}

\begin{center}
  \bf  A. V. Latyshev\footnote{$avlatyshev@mail.ru$} and
  A. A. Yushkanov\footnote{$yushkanov@inbox.ru$}
\end{center}\medskip

\begin{center}
{\it Faculty of Physics and Mathematics,\\ Moscow State Regional
University,  105005,\\ Moscow, Radio st., 10--A}
\end{center}\medskip

\begin{abstract}
It is shown that for thin metallic films thickness of which does not
exceed thickness of skin layer, the problem allows analytical
solution. In the field of resonant  frequencies the analysis of
dependence of coefficients of transmission, reflection and
absorbtion on an electromagnetic wave is carried out. Dependence on
pitch angle, thickness of the layer and coefficient of specular
reflection and on effective electron collision frequency is carried
out. The formula for contactless determination (calculation) of a
thickness of a film by observable resonant frequencies is deduced.
\medskip

\medskip

structures
\end{abstract}

\begin{center}\bf
  1. Introduction
\end{center}

The problem of interaction of an electromagnetic wave with the metal
film attracts attention to itself for the long time already
\cite{F69} -- \cite{Landau8}. It is connected as with theoretical
interest to this problem, and with numerous practical appendices as
well \cite{Antonez} and \cite{A2004}.

Researches of interaction of an electromagnetic wave with metallic films
were carried out basically for a
case of specular dissipation of electrons on a film surface. It is
connected with the fact that for more general boundary conditions
the problem becomes essentially complicated and does not suppose the
analytical solution generally.

In the present work it is shown that for thin films, a thickness of
which does not exceed a thickness of a skin -- layer, the problem
allows the analytical solution. In previous work \cite{LY2010}  the
case when the frequency of electromagnetic wave is less than plasma
(Langmuir) frequency was considered.

Let us note, that the most part of reasonings carried out below
would be true for the more general case of conducting (in
particular, semi-conductor) film.\medskip

\begin{center}
{\bf 2. Problem Statement}
\end{center}

We consider the thin slab of conducting medium on which the
elec\-tro\-mag\-netic wave falls. We denote the pitch angle by
$\theta$. We will assume, that the vector of magnetic field of
electromagnetic wave is parallel to the surface of the slab. Such
wave is called $p$--wave (see \cite{K} or \cite{F69}).

We take Cartesian coordinate system with origin of coordinates on
one of the surfaces of a slab, with axis $x$, directed deep into the
slab. We direct the axis $y$ parallel with the vector of magnetic
field of electromagnetic wave.
Under such choice of the system of coordinates the 
electric field vector and magnetic field vector have the
following structure

$\mathbf{E}=\{E_x(x,y,z),0,E_z(x,y,z)\}$ and
$\mathbf{H}=\{0,H_y(x,y,z),0\}$.

Further components of electric and magnetic field vectors
we search in the form
$$
E_x(x,z,t)=E_x(x)e^{-i\omega t+ik\sin \theta z},\quad
E_z(x,z,t)=E_z(x)e^{-i\omega t+ik\sin \theta z},
$$
and
$$
H_y(x,z,t)=H_y(x)e^{-i\omega t+ik\sin \theta z}.
$$

Now behaviour of electric and magnetic fields of the wave in the
slab is described by the following system of differential equations
\cite{K}
$$
\left\{\begin{array}{l}
\dfrac{dE_z}{dx}-ik\sin\theta E_x+ikH_y=0 \\ \\
ikE_x-ik\sin\theta H_y=\dfrac{4\pi}{c}j_x\\ \\
\dfrac{dH_y}{dx}+ikE_z=\dfrac{4\pi}{c}j_z.
\end{array}\right.
\eqno{(1)}
$$

Here $c$ is the velocity of light, $\,\bf j$ is the current density,
$k$ is the wave number.

We denote the thickness of the slab by $d$.

The coefficients of transmission $T$, reflection $R$ and absorption
$A$ of the electromagnetic wave by the slab are described by the
following expressions \cite{F69}, \cite{F66}
$$
T=\dfrac{1}{4}\big|P^{(1)}-P^{(2)}\big|^2,
\eqno{(2a)}
$$

$$
R=\dfrac{1}{4}\big|P^{(1)}+P^{(2)}\big|^2,
\eqno{(2b)}
$$
and
$$
A=1-T-R.
\eqno{(2c)}
$$

The quantities $P^{(j)}\;(j=1,2)$ are defined by the following
expressions
$$
P^{(j)}=\dfrac{\cos \theta+Z^{(j)}}{\cos \theta-Z^{(j)}},\qquad j=1,2.
\eqno{(3)}
$$

The quantity $Z^{(1)}$ corresponds to impedance on the lower surface
of slab under symmetrical configuration of the external magnetic
field. This is the case 1 when
$$
H_y(0)=H_y(d),\qquad  E_x(0)=E_x(d),\qquad E_z(0)=-E_z(d)
$$
The quantity $Z^{(2)}$ corresponds to the impedance on the lower
surface of slab under antysymmetrical configuration of the external
magnetic field. This is the case 2 when
$$
H_y(0)=-H_y(d),\qquad  E_x(0)=-E_x(d),\qquad
E_z(0)=E_z(d).
$$

The impedance is thus defined as follows

$$
Z^{(j)}=\dfrac{E_z(-0)}{H_y(-0)}, \qquad j=1,2.
\eqno{(4)}
$$

We will consider the case when the width of the slab $d$ is less
than the depth of the skin -- layer $\delta$. Let's note, that depth
of the skin -- layer depends essentially on frequency of radiation,
monotonously decreasing in process of growth of the last. The value
$\delta$ possesses the minimum value in so-called infrared case
\cite{Landau10}
$$
\delta_0=\dfrac{c}{\omega_p},
$$
where $\omega_p$ is the plasma frequency.

For typical metals  \cite{Landau10} $\delta_0\sim 100$ nm.

Hence for the films thickness of which $d$ is less than $\delta_0$
our assumption is true for any frequencies.

\begin{center}\bf
3.  Problem solution
\end{center}

The quantities $H_y $ and $E_z $ change a little on distances
smaller than the depth of a skin -- layer.

Therefore under fulfilment of the given assumption $d<\delta$ these
electrical and magnetical fields will change a little in the slab.

In case 1 when $H_y(0)=H_y(d)$ it is possible to accept that the
value $H_y$ is constant within the slab.

Variation of the quantity of $y$-projection of electric field on
thickness of slab can be defined from the first equation of the
system (1)
$$
E_y(d)-E_y(0)=-ikd H_y+ik\sin\theta \intl_0^d E_x dx.
\eqno{(5)}
$$

From  the second equation of the system (1) it follows that on the
boundary of the film the following relationship is satisfied
$$
E_x(0)=E_x(d)=H_y\sin\theta.
\eqno{(6)}
$$

The integral from the relation (5) is proportional to the value of
the quantity of normal to the surface component of electrical field
on the surface and therefore according to the relation (6) it's
proportional to the quantity $H_y$.

We define the coefficient of proportionality as
$$
G=\dfrac{1}{{E_x(0)d}}\int\limits_{0}^{d}E_x(x)\,dx=
\dfrac{1}{H_y d\sin \theta}\int\limits_{0}^{d}E_x(x)\,dx.
\eqno{(7)}
$$

With the help of (7) we rewrite the relation (5) in the following
form
$$
E_z(d)-E_z(0)=\big(-ikd +ikGd\sin^2\theta\big)H_y.
$$

Considering the antisymmetric character of the projection of
electric field $E_y$ in this case we receive
$$
E_y(0)=ikd\big(1-G\sin^2\theta\big)\dfrac{H_y}{2}.
\eqno{(8)}
$$

Therefore for the impedance we have
$$
Z^{(1)}=\dfrac{ikd}{2}\big(1 -G\sin^2\theta\big).
\eqno{(9)}
$$

For the case 2 when $E_z(0)=E_z(d)$, it is possible to assume that
$z$ -- projection of electric field $E_z$ is constant in the slab.

Then the magnetic field change on the width of a slab can be
determined from the third equation of the system (1)
$$
H_y(d)-H_y(0)=-ikdE_z+\dfrac{4\pi}{c}\intl_0^d j_z(x)dx.
\eqno{(10)}
$$

Thus
$$
j_z(x)=\sigma(x)E_z,
$$
where $\sigma(x)$ is the conductance that in general case depends
on coordinate $x$.

Let's introduce the longitudinal conductivity averaged by thickness
of the slab,

$$
\sigma_d=\dfrac{1}{E_zd}\intl_0^d j_z(x) dx=\dfrac{1}{d}
\intl_0^d \sigma(x) dx.
\eqno{(11)}
$$

Then the relation (10) can be rewritten with help (11) in the following form
$$
H_y(d)-H_y(0)=-ikd E_z+\dfrac{4\pi d\sigma_d}{c}E_z.
$$

Considering sym\-met\-ry of the magnetic field, from here we have
$$
H_y(0)=\dfrac{1}{2}ikdE_z-\dfrac{2\pi d \sigma_d}{c}E_z.
$$

For the impedance (4) we have

$$
Z^{(2)}=\dfrac{2c}{ickd-4\pi \sigma_d d }.
\eqno{(12)}
$$

\begin{center}\bf
4.  Coefficients of transmission, reflection and absorption
\end{center}

From here according to (3) we receive expressions for the quantities
$P^{(j)}\;(j=1,2)$

$$
P^{(1)}=\dfrac{2\cos \theta+ikd(1-G\sin^2\theta)}
{2\cos \theta-ikd(1-G\sin^2\theta)},
\eqno{(13a)}
$$
$$
\quad P^{(2)}=\dfrac{(4\pi\sigma_d -ikc) d\cos \theta-2c}
{(4\pi\sigma_d -ikc) d\cos \theta+2c}.
\eqno{(13b)}
$$

We will assume that length of a wave of incident radiation surpasses
essentially the thickness of the slab, i.e. $kd\ll 1$. Then
expressions (9) and (12) for impedances and expression (13) for
quantity $P^{(j)}\;(j=1,2)$ become a little simpler
$$
Z^{(1)}_0=-\dfrac{1}{2}ikGd\sin^2\theta,
 \qquad Z^{(2)}_0=-\dfrac{c}{2\pi \sigma_d d }.
\eqno{(14)}
$$
Substituting (14) into (3), we have:
$$
P^{(1)}_0=
\dfrac{2\cos \theta-ikGd\sin^2\theta}{2\cos \theta+ikGd\sin^2\theta},
\qquad
P^{(2)}_0=
\dfrac{2\pi \sigma_d d\cos \theta-c}{2\pi \sigma_d d\cos \theta+c}.
\eqno{(15)}
$$

Thus quantities $R, T, A $ can be found according to the formulas
(2).

In a limiting case of non-conducting slab, when $\sigma_d\to 0$,
$G\to 1$ from these expressions we have
$$
P^{(1)}=-P^{(2)}=\dfrac{2+ikd\cos \theta}{2-ikd\cos \theta},
$$
from which
$$
T= 1,\qquad \, R= 0, \,\qquad A=0.
$$

Under almost tangent incidence when $\theta\to \pi/2$ we receive
$P^{(1)}\to -1,P^{(2)}\to -1$. Thus, we obtain that $T\to 0,\, R\to
1,\,A\to 0$.

Let the relation $kl\ll 1$ be true. Then in a low-frequency case,
when $\omega\to 0$, the quantity $\sigma_d$ for a metal film can be
presented in the following form \cite {S}
$$
\sigma_d=\dfrac{w}{\Phi(w)}\,\sigma_0,\quad\quad
w=\dfrac{d}{l},
\eqno{(16)}
$$
where
$$
\dfrac{1}{\Phi(w)}=\dfrac{1}{w}-\dfrac{3}{2w^2}(1-p)\intl_1^\infty\Big(
\dfrac{1}{t^3}-\dfrac{1}{t^5}\Big)\dfrac{1-e^{-wt}}{1-pe^{-wt}}dt.
$$

Here $l$ is the mean free path of electrons, $p$ is the coefficient
of specular reflection, $\sigma_0=\omega_p^2\tau/(4\pi)$ is the
static conductivity of a volume pattern, $\tau=l/v_F$ is the time of
mean free path of electrons, $v_F$ is the Fermi velocity.

In a low-frequency case when the formula (16) is applicable, the
coefficients $T, R, A $ do not depend on frequency of the incoming
radiation according to the formulas (2).

For arbitrary frequencies these expressions are true under
condition, that it is necessary to use the following expression $
l\to \dfrac{v_F \tau}{1-i\omega\tau}, $ as a quantity $l$ and the
expression $ \sigma_0\to \dfrac{\sigma_0}{1-i\omega\tau}$ instead of
$\sigma_0$.

For the case $kl\ll 1$ the quantity $G$ can be calculated from the
problem of behaviour of a plasma slab in variable electric field,
which is perpendicular to the surface of slab \cite {LY2008}.

Let's calculate the coefficients of transmission and reflection in
the case when $kd\ll 1$. We will substitute expressions $P^{(1)}$
and $P^{(2)}$, defined according to (15), into the formulas (2).
We receive
$$
T=\cos^2\theta\Bigg|\dfrac{1-ik\dfrac{d}{2}G\sin^2\theta
\dfrac{2\pi d\sigma_d}{c}}{(\cos\theta+ik\dfrac{d}{2}G
\sin^2\theta)(1+\dfrac{2\pi d\sigma_d}{c}\cos\theta)}\Bigg|^2,
\eqno{(17)}
$$
$$
R=\Bigg|\dfrac{ik\dfrac{d}{2}G\sin^2\theta-
\dfrac{2\pi d\sigma_d}{c}\cos^2\theta}{(\cos\theta+ik\dfrac{d}{2}G
\sin^2\theta)(1+\dfrac{2\pi d\sigma_d}{c}\cos\theta)}\Bigg|^2,
\eqno{(18)}
$$
where the quantity $G$ may be found from the solution of the problem
of plasma oscillations  \cite{Les} and \cite {LY2008},
$$
G=\dfrac{1}{d}\int\limits_{0}^{d}e(x)dx,
\eqno {(19)}
$$
and $e(x)$ is the electric field.

\begin{center}\bf 5.
Coefficients of transmission, reflection and absorption in the case
of specular reflection of electrons
\end{center}

In the case of mirror reflection of electrons according to (16)  we
have $\sigma_d=\sigma_0$, and the quantity $G$ under the formula
(19) can be calculated precisely, using the electric field in plasma
layer constructed in \cite {LY2008} .

At the proof of decomposition of the decision of an initial boundary
problem in \cite {LY2008}  the electric field in a metal layer has
actually been constructed
$$
e(x)=\dfrac{\lambda_1}{\lambda_\infty}+
\dfrac{2\lambda_1\eta_0\cosh(z_0x/\eta_0)}
{(ac-\eta_0^2)\lambda'(\eta_0) \cosh(z_0/\eta_0)}+
$$
$$\qquad +\dfrac{\lambda_1}{2}\int\limits_{-1}^{1}
\dfrac{\eta^2\cosh(z_0x/\eta)}{\lambda^+(\eta)\lambda^-(\eta)\cosh(z_0/
\eta)}\,d\eta.
$$

In this formula $\lambda(z)$ is the dispersion function from the
problem of plasma oscillations,
$$
\lambda(z)=c^2+\dfrac{z^2}{2}\int\limits_{-1}^{1}\dfrac{\eta_1^2-
\tau^2}{\tau^2-z^2}d\tau, \qquad \eta_1^2=ac, \qquad a=\dfrac{a_0\nu}
{v_F\varkappa},
$$
$$
\varkappa=\dfrac{9a_0^2}{r_D^2},\qquad
r_D^2=\dfrac{3v_F^2}{\omega_p^2},\qquad c= \dfrac{a_0(\nu -i
\omega)}{\varkappa v_F},\qquad z_0=c\varkappa,
$$
$r_D$ is the Debaye radius, $\lambda_1=\lambda(\eta_1)=c^2-ac$,
$\eta_k^*$ are the zeroes of the functions $\cosh(z_0/\eta)$,
$$
\eta_k^*=-\dfrac{2z_0i}{\pi(2k+1)}=-\dfrac{2a_0(\omega+i \nu)}
{\pi(2k+1)v_F}, \qquad k=0,\pm 1,\pm . 2,\cdots
$$

In the layer $0\leq x\leq d$ the electric field has the following
form
$$
e(x)=\dfrac{\cosh[z_0(2x-d)/\eta_1 d]}{\cosh(z_0/\eta_1)}-
\dfrac{\lambda_1}{z_0}
\sum\limits_{k=-\infty}^{k=+\infty}\dfrac{{\eta_k^*}^3
\cosh[z_0(2x-d)/\eta_k^*d]}
{\lambda(\eta_k^*)({\eta_k^*}^2-\eta_1^2)\sinh(z_0/\eta_k^*)}.
\eqno{(20)}
$$

The quantity $G$ can be found easily by means of the equalities (19)
and (20) (see also \cite {LY2008}) and has the following form:
$$
G=\dfrac{\lambda_1}
{\lambda_\infty}+\dfrac{2\lambda_1\eta_0^2\tanh(z_0/\eta_0)}
{z_0(ac-\eta_0^2)\lambda'(\eta_0)} +\dfrac{\lambda_1}{2z_0}
\int\limits_{-1}^{1}\dfrac{\tanh(z_0/\eta)\eta^3\,d\eta}
{\lambda^+(\eta)\lambda^-(\eta)}.
\eqno{(21)}
$$

Here
$$
\lambda^{\pm}(\mu)=\lambda(\mu)\pm
i\dfrac{\pi}{2}\mu(\eta_1^2-\mu^2).
$$

Integral from (21) we will calculate by means of methods of contour
integration. Let us take advantage further of obvious equality
$$
\dfrac{1}{\lambda^+(\mu)\lambda^-(\mu)}= \dfrac{1}{i\pi\mu(\mu^2-
\eta_1^2)}\Bigg[\dfrac{1}{\lambda^+(\mu)}-\dfrac{1}{\lambda^+(\mu)}\Bigg].
$$

Therefore the integral from (21) is equal to
$$
\dfrac{1}{2}\int\limits_{-1}^{1}\dfrac{\tanh(z_0/\tau)\tau^3\,d\tau}
{\lambda^+(\tau)\lambda^-(\tau)}=\dfrac{1}{2\pi i}
\int\limits_{-1}^{1}\Bigg[\dfrac{1}{\lambda^+(\tau)}-\dfrac{1}
{\lambda^-(\tau)}\Bigg]\dfrac{\tanh(z_0/\tau)\tau^2\,d\tau}
{\tau^2-\eta_1^2}.
$$

We take now a circle $\gamma_R$ ($\gamma_R: |z|=R)$  with so big
radius $R$, that all finite singular points of function
$$
f(z)=\dfrac{\tanh(z_0/z)z^2}{(z^2-\eta_1^2)\lambda(z)}
$$
lay inside $\gamma_R$. Such points are points $z=\pm \eta_1$, zeroes
of the dispersion function $\lambda(z)$ are points $z=\pm \eta_0$
(if $(\gamma, \varepsilon)\in D^+$) (see \cite {LY2008}), and also
polar singularity of the function $\tanh(z_0/z)$. The last points
are points $z=\eta_k^*,\;k=0,\pm1,\pm2,\cdots $.

According to the theorem of the full sum of residues we receive
$$
\Bigg[\Res_{z=\infty}+\sum\limits_{k=-\infty}^{k=+\infty}\Res_{z=\eta_k^*}
+\Res_{z=\eta_0}+\Res_{z=-\eta_0}+\Res_{z=\eta_1}+\Res_{z=-\eta_1}\Bigg]
\dfrac{\tanh(z_0/z)z^2}{(z^2-\eta_1^2)\lambda(z)}=
$$
$$
=\dfrac{1}{2\pi i}
\int\limits_{-1}^{1}\Bigg[\dfrac{1}{\lambda^+(\tau)}-\dfrac{1}
{\lambda^-(\tau)}\Bigg]\dfrac{\tanh(z_0/\tau)\tau^2\,d\tau}
{\tau^2-\eta_1^2}.
$$

We note that
$$
\Res_{z=\infty}\dfrac{\tanh(z_0/z)z^2}{(z^2-\eta_1^2)\lambda(z)}=-
\dfrac{z_0}{\lambda_\infty},
$$
$$
\Res_{z=\eta_k^*} \dfrac{\tanh(z_0/z)z^2}{(z^2-\eta_1^2)\lambda(z)}=-
\dfrac{2{\eta_k^*}^4}{z_0\lambda(\eta_k^*)({\eta_k^*}^2-\eta_1^2)},
$$
$$
\Res_{z=\pm \eta_1} \dfrac{\tanh(z_0/z)z^2}{(z^2-\eta_1^2)\lambda(z)}=
\dfrac{\eta_1\tanh(z_0/\eta_1)}{2\lambda(\eta_1)},
$$
$$
\Res_{z=\pm \eta_0} \dfrac{\tanh(z_0/z)z^2}{(z^2-\eta_1^2)\lambda(z)}=
\dfrac{\eta_0^2\tanh(z_0/\eta_0)}{\lambda'(\eta_0)(\eta_0^2-\eta_1^2)}.
$$

Thus, the integral from (21) is equal to
$$
\dfrac{1}{2}\int\limits_{-1}^{1}\dfrac{\tanh(z_0/\tau)\tau^3\,d\tau}
{\lambda^+(\tau)\lambda^-(\tau)}=-\dfrac{z_0}{\lambda_\infty}+
\dfrac{\eta_1\tanh(z_0/\eta_1)}{2\lambda(\eta_1)}+
$$
$$
+\dfrac{2\eta_0^2\tanh(z_0/\eta_0)}{\lambda'(\eta_0)(\eta_0^2-\eta_1^2)}-
2\sum\limits_{n=-\infty}^{+\infty}
\dfrac{{\eta_n^*}^4}{\lambda(\eta_n^*)({\eta_n^*}^2-\eta_1^2)}.
$$

Hence, the quantity $G$ is equal to
$$
G=\dfrac{\eta_1}{z_0}\tanh\dfrac{z_0}{\eta_1}- \dfrac{\lambda_1}{z_0^2}
\sum\limits_{n=-\infty}^{+\infty}
\dfrac{{\eta_n^*}^4}{\lambda(\eta_n^*)({\eta_n^*}^2-\eta_1^2)}.
\eqno{(22)}
$$

Let us note, that according to the evenness of the expression
standing under a sign of the sum in (22),  it is possible to
simplify this sum and present the expression (22) in the following
form
$$
G=\dfrac{\eta_1}{z_0}\tanh\dfrac{z_0}{\eta_1}-\dfrac{2\lambda_1}{z_0^2}
\sum\limits_{k=0}^{k=+\infty}\dfrac{{\eta_k^*}^4}{
\lambda(\eta_k^*)({\eta_k^*}^2-\eta_1^2)}. \eqno{(23)}
$$

Let us note, that the sum of the series from (23) is well
approximated by the first member, i.e. instead of (23) it is
possible to take
$$
G_1=\dfrac{\eta_1}{z_0}\tanh\dfrac{z_0}{\eta_1}-
\dfrac{2\lambda_1{\eta_0^*}^4}{z_0^2
\lambda(\eta_0^*)({\eta_0^*}^2-\eta_1^2)}.
$$

Let us carry out numerical calculations. We will enter the relative
error
$$
O_1(\Omega,\varepsilon,d)=\left|\dfrac{G-G_1}{G}\right|\cdot
100\%,\quad \Omega=\dfrac{\omega}{\omega_p}, \quad
\varepsilon=\dfrac{\nu}{\omega_p}.
$$

Then for a film of sodium ($\omega_p=6.5\cdot 10^{15} \sec^{-1}$,
$v_F=8.52\cdot 10^7$ cm/sec) of the thickness of 1 nanometer, 5
nanometers and 10 nanometers under $\omega=\omega_p$ and
$\nu=10^{-3}\omega_p \sec^{-1}$ accordingly we have: $O_1=1.42 \%,
1.38 \% $ and $1.98 \% $. For $G$ we have replaced an infinite
series with the finite sum for $N=10^6$ members.

The quantity $G$ is approximated by first two members of the
decomposition (21) even more effectively, i.e. when  we replace $G$
with the quantity
$$
G_2=\dfrac{\lambda_1}
{\lambda_\infty}+\dfrac{2\lambda_1\eta_0^2\tanh(z_0/\eta_0)}
{z_0(ac-\eta_0^2)\lambda'(\eta_0)}.
\eqno{(24)}
$$

The formula (24) means, that we have replaced the electric field
(20) by two first components of Drude and Debaye, corresponding to
the discrete spectrum.

For the calculation of the quantity $G_2$  explicit expression of
zero of dispersion function $\eta_0 =\eta_0 (\Omega, \varepsilon)$
is required. We will write  the factorization formula of dispersion
function (see \cite{LY2008}) without the proof
$$
\lambda(z)=\lambda_\infty(\eta_0^2-z^2)X(z)X(-z).
\eqno{(25)}
$$

In (25) we introduce the following notations
$$
\lambda_\infty=\lambda(\infty)=\dfrac{1}{3}+ac-c^2=\dfrac{1}{3}(1-\Omega^2-
i\varepsilon \Omega),
$$
$$
X(z)=\dfrac{1}{z}\exp V(z), \quad V(z)=\dfrac{1}{2\pi
i}\int\limits_{0}^{1}\dfrac{\ln G(\tau)-2\pi i} {\tau-z}d\tau,
$$
$$
G(\tau)=\dfrac{\lambda^+(\tau)}{\lambda^-(\tau)},\quad
\lambda^{\pm}(\tau)=c^2-ac-(\tau^2-ac)\lambda_0^{\pm}(\tau),
$$
$$
\lambda_0^{\pm}(\tau)=\lambda_0(\tau)\pm \dfrac{\pi}{2}\tau i,\quad
\lambda_0(\tau)=1+\dfrac{\tau}{2}
\int\limits_{-1}^{1}\dfrac{d\tau'}{\tau'-\tau}=
1+\dfrac{\tau}{2}\ln\dfrac{1-\tau}{1+\tau}.
$$

If we calculate values of the left and right parts of the equation
(25) in the point $z=i $, then for square of zero of dispersion
function after some transformations we receive following expression
$$
\eta_0^2=-1+\dfrac{\lambda(i)}{\lambda_\infty X(i)X(-i)}=
-1+\dfrac{\lambda(i)}{\lambda_\infty}\exp\Big[-V(i)-V(-i)\Big].
$$

Considering, that $ \lambda_0(i)=1-\dfrac{\pi}{4}$, we have
$$
\lambda(i)=c^2-ac+\Big(1-\dfrac{\pi}{4}\Big)(1+ac)=
-\dfrac{1}{3}(\Omega^2+i\varepsilon\Omega)+\Big(1-\dfrac{\pi}{4}\Big)
\Big[1+\dfrac{1}{3}(\varepsilon^2-i\varepsilon\Omega)\Big].
$$

It is possible to present the function $X(z)=\dfrac{1}{z}e^{V (z)}$
in the following form
$$
X(z)=\dfrac{1}{z-1}e^{V_0(z)}, \qquad V_0(z)=\dfrac{1}{2\pi i}
\int\limits_{0}^{1}\dfrac{\ln G(\tau)d\tau}{\tau-z}.
$$

Let us find the sum
$$
V_0(i)+V_0(-i)=\dfrac{1}{2\pi i}\int\limits_{0}^{1}\dfrac{\ln
G(\tau)d\tau} {\tau-i}+\dfrac{1}{2\pi i}\int\limits_{0}^{1}\dfrac{\ln
G(\tau)d\tau} {\tau+i}=
$$
$$
=\dfrac{1}{2\pi i}\int\limits_{-1}^{1} \dfrac{\ln
G(\tau)d\tau}{\tau-i}=\dfrac{1}{2\pi i}\int\limits_{-1}^{1}
\dfrac{\tau\ln G(\tau)d\tau}{\tau^2+1}.
$$

By means of these formulas we will transform the formula for the
square of zero of the dispersion functions to the form
$$
\eta_0^2=-1+\dfrac{2\lambda(i)}{\lambda_\infty}\exp\Bigg[
-\dfrac{1}{2\pi i}\int\limits_{-1}^{1}\dfrac{\tau G_l(\tau)d\tau}
{\tau^2+1}\Bigg],
$$
or
$$
\eta_0^2=-1+\dfrac{2\lambda(i)}{\lambda_\infty}\exp\Bigg[
\dfrac{i}{\pi}\int\limits_{0}^{1}\dfrac{\tau G_l(\tau)d\tau}
{\tau^2+1}\Bigg], \eqno{(26)}
$$
where
$$
G_l(\tau)=\ln\dfrac{(3\tau^2-\varepsilon^2+i\varepsilon\Omega)
(\lambda_0(\tau)+\dfrac{\pi}{2}\tau i)+\Omega^2+i\varepsilon\Omega}
{(3\tau^2-\varepsilon^2+i\varepsilon\Omega)
(\lambda_0(\tau)-\dfrac{\pi}{2}\tau i)+\Omega^2+i\varepsilon\Omega}.
$$

Now a relative error
$$
O_2(\Omega,\varepsilon,d)=\left|\dfrac{G-G_2}{G}\right|\cdot 100 \%
$$
for sodium films of the thickness of 1 nanometer, 5 nanometers and
10 nanometers at $\omega =\omega_p$ and $\nu=10^{-3}\omega_p$
accordingly equals: $O_2=0.575 \%, 0.003 \% $ and $0.0004 \% $.

The graph of the relative error $O_2(\Omega,\varepsilon,d)$ as a
function of the variable $\Omega$ for a film with the thickness of
$10$ nm at $\nu=0.001\omega_p \sec^{-1}$ is represented on the Fig.
1.

\begin{figure}[h]\center
\includegraphics[width=14.0cm, height=7.5cm]{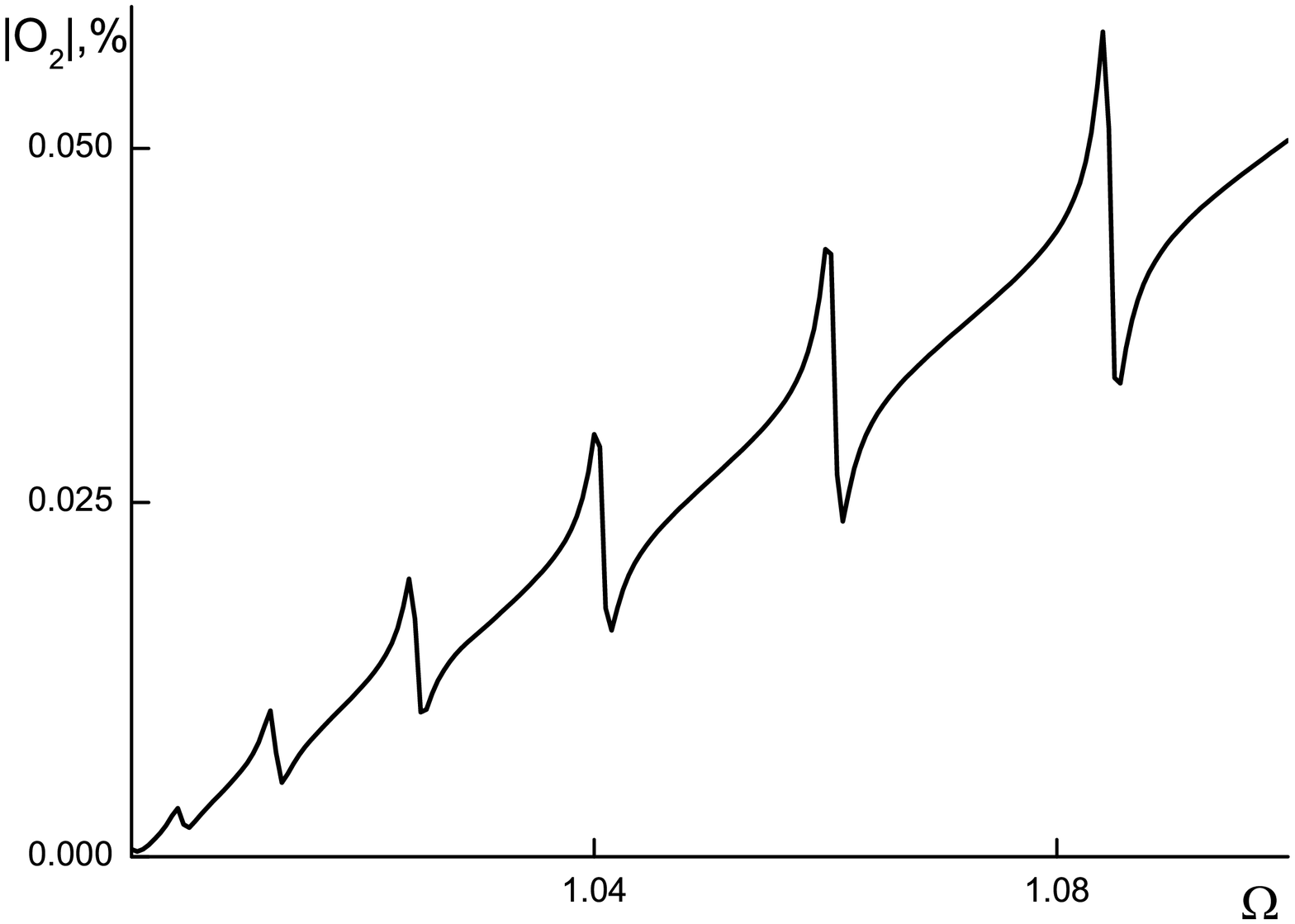}
\noindent\caption{Influence of continuous spectrum, $d=10$ nm,
$\nu=0.001\omega_p$, $\theta=75^\circ$.}\end{figure}

The graph on Fig. 1 shows, that in the area $\omega\geqslant
\omega_p $ the contribution to the electric field, corresponding to
the continuous spectrum, is insignificant, and can be neglected.
Thus, function $G$ is approximated by two composed Drude and Debaye,
corresponding to discrete spectrum, according to (24).

Using the formulas (17) and (18), with the use of the expression
(24) for functions $G $, we will carry out graphic research of the
coefficients of transmission, reflecion and absorption.

\begin{center}\bf
6.  Discussion of the results
\end{center}

Let us consider the case of thin sodium film. We will construct
graphics of the dependences of transmission, reflecion and
absorption on quantity $\Omega =\omega/\omega_p $ at pitch angle
 $ \theta=75 ^\circ $ (Figs. 2 -- 10).

We will note, that near to a plasma resonance ($\omega\sim
\omega_p$) the coefficient of transmissiom has a minimum, and
reflecion and absorption coefficients have minimum values. Under a
thickness of the film of 1.5 nanometers and at $\nu=0.05\omega_p$
in the areas of resonant frequencies ($\omega>\omega_p $)
all coefficients have one maximum more. Under increase of the
thickness of the film from 1.5 to 10 nanometers the second maximum
vanishes.

For a film with the thickness of 5 nanometers and under $ \nu=0.02
\omega_p$ in the field of resonant frequencies behaviour
of all coefficients has so-called "edge" \, character ("paling").
Under the further increase of the  thickness of the film frequency
of "combs" \, increases also, and we can see growth of quantity of
its "teeths" \, (Figs. 5 -- 10). If for a film with the thickness of
5 nanometers the quantity $\varepsilon=\nu/\omega_p $ decreases, the
quantity of "teeths" \, of combs grows sharply.

\begin{figure}[t]\center
\includegraphics[width=14.0cm, height=6cm]{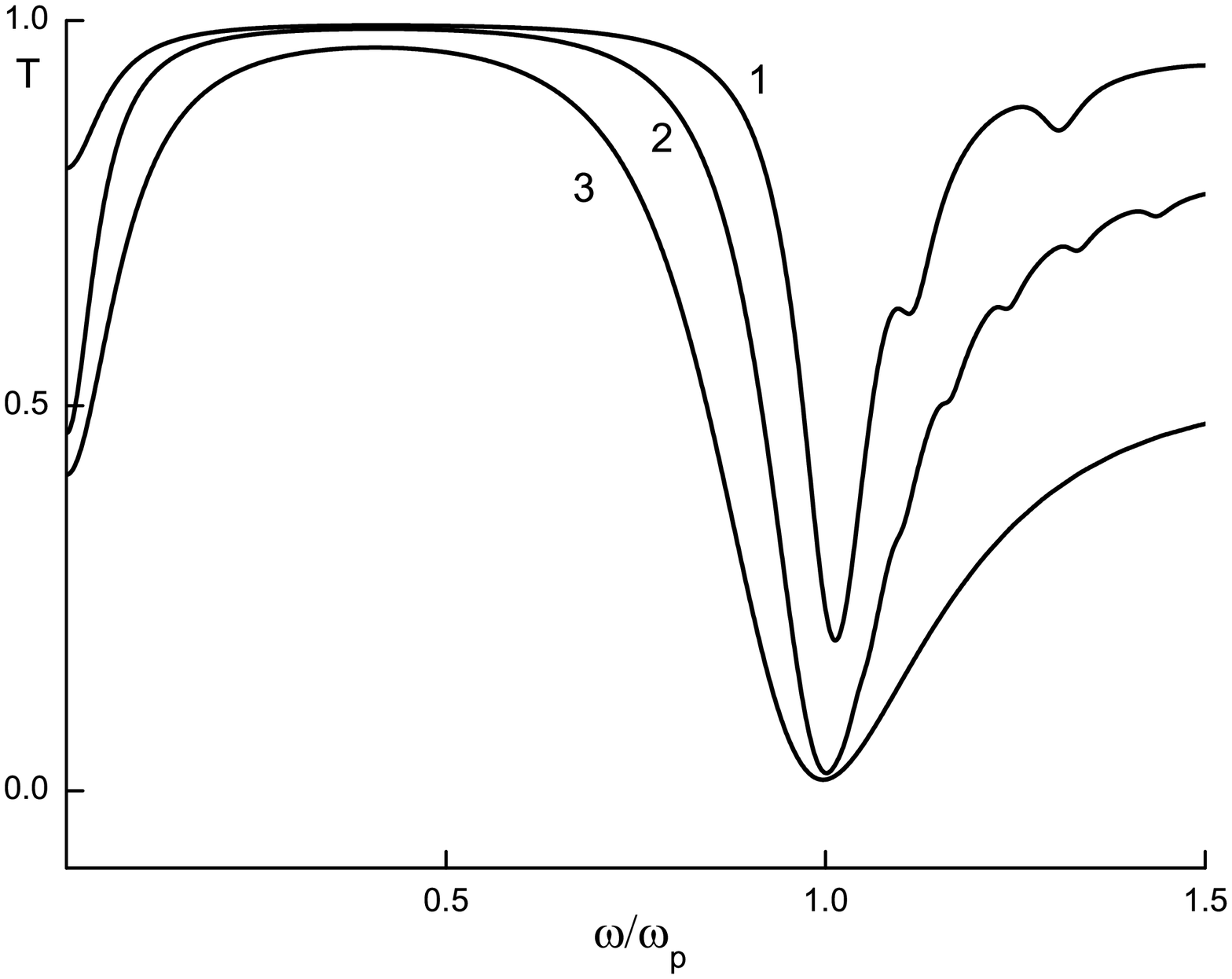}
\noindent\caption{Transmittance, $\theta=75^\circ$. Curves
$1,2,3$ correspond to values of parameters $d=2$ nm,
$\nu=0.05\omega_p$; $d=5$ nm, $\nu=0.03\omega_p$;\;
$d=10$ nm,\; $\nu=0.05\omega_p$.}
\includegraphics[width=14.0cm, height=6cm]{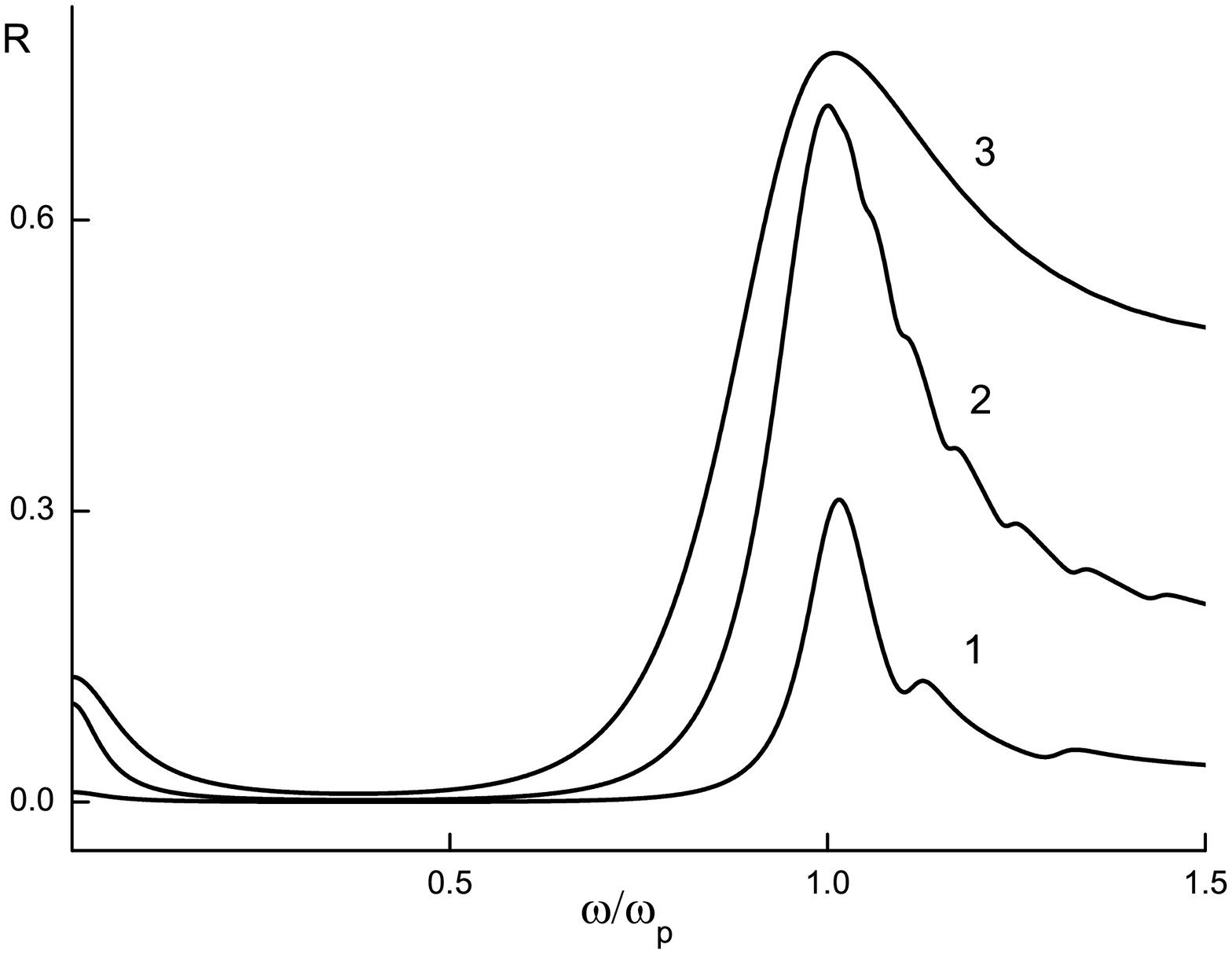}
\noindent\caption{Reflectance, $\theta=75^\circ$. Curves
$1,2,3$ correspond to values of parameters $d=2$ nm,
$\nu=0.05\omega_p$; $d=5$ nm,
$\nu=0.03\omega_p$;
$d=10$ nm, $\nu=0.05\omega_p$.}
\includegraphics[width=14.0cm, height=7cm]{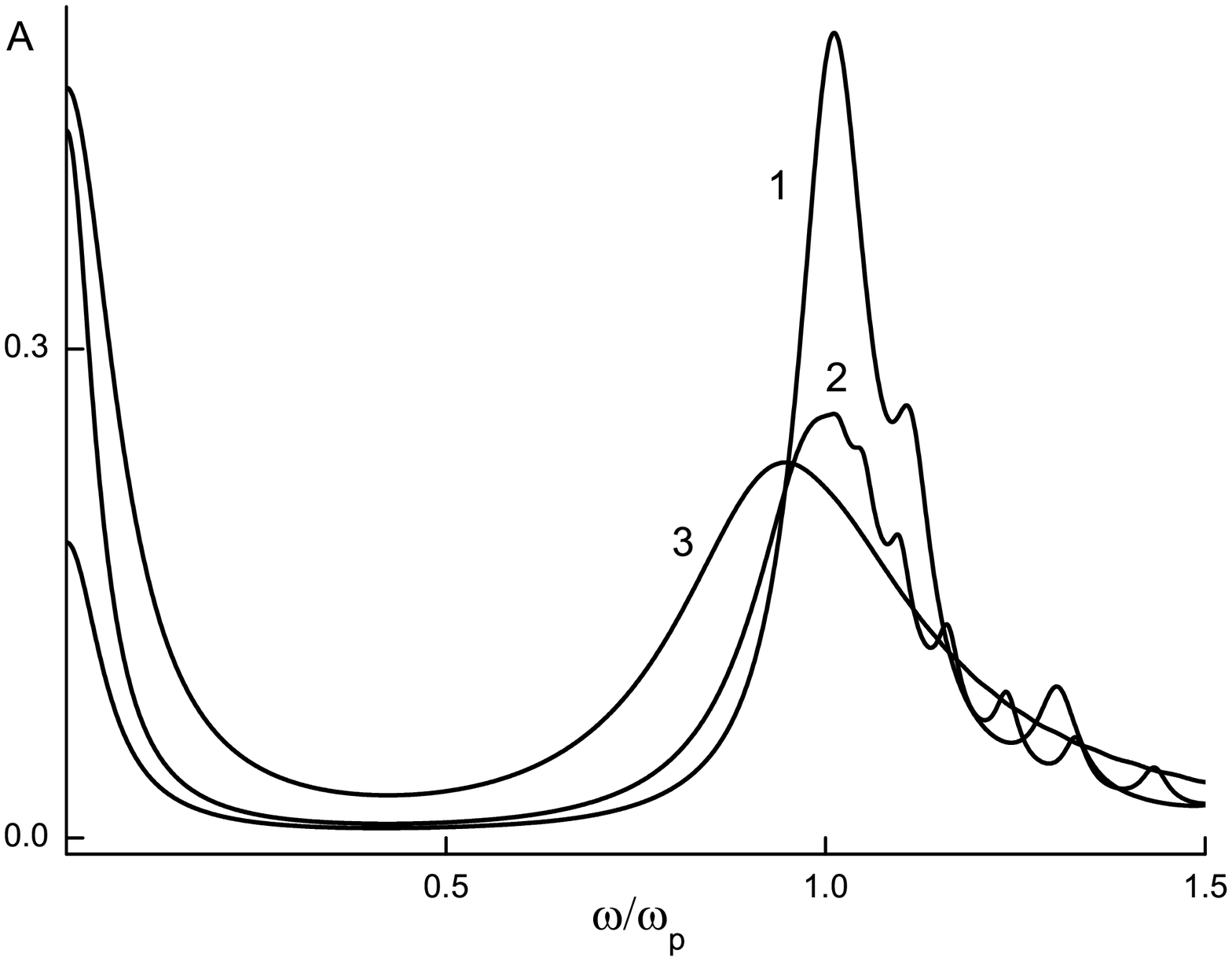}
\noindent\caption{Absorptance, $\theta=75^\circ$.
Curves $1,2,3$ correspond to values of parameters $d=2$ nm,
$\nu=0.05\omega_p\; \sec^{-1}$; $d=5$ nm, $\nu=0.03\omega_p$,
$d=10$ nm, $\nu=0.05\omega_p$.}
\end{figure}

\begin{figure}[t]\center
\includegraphics[width=16.0cm, height=6cm]{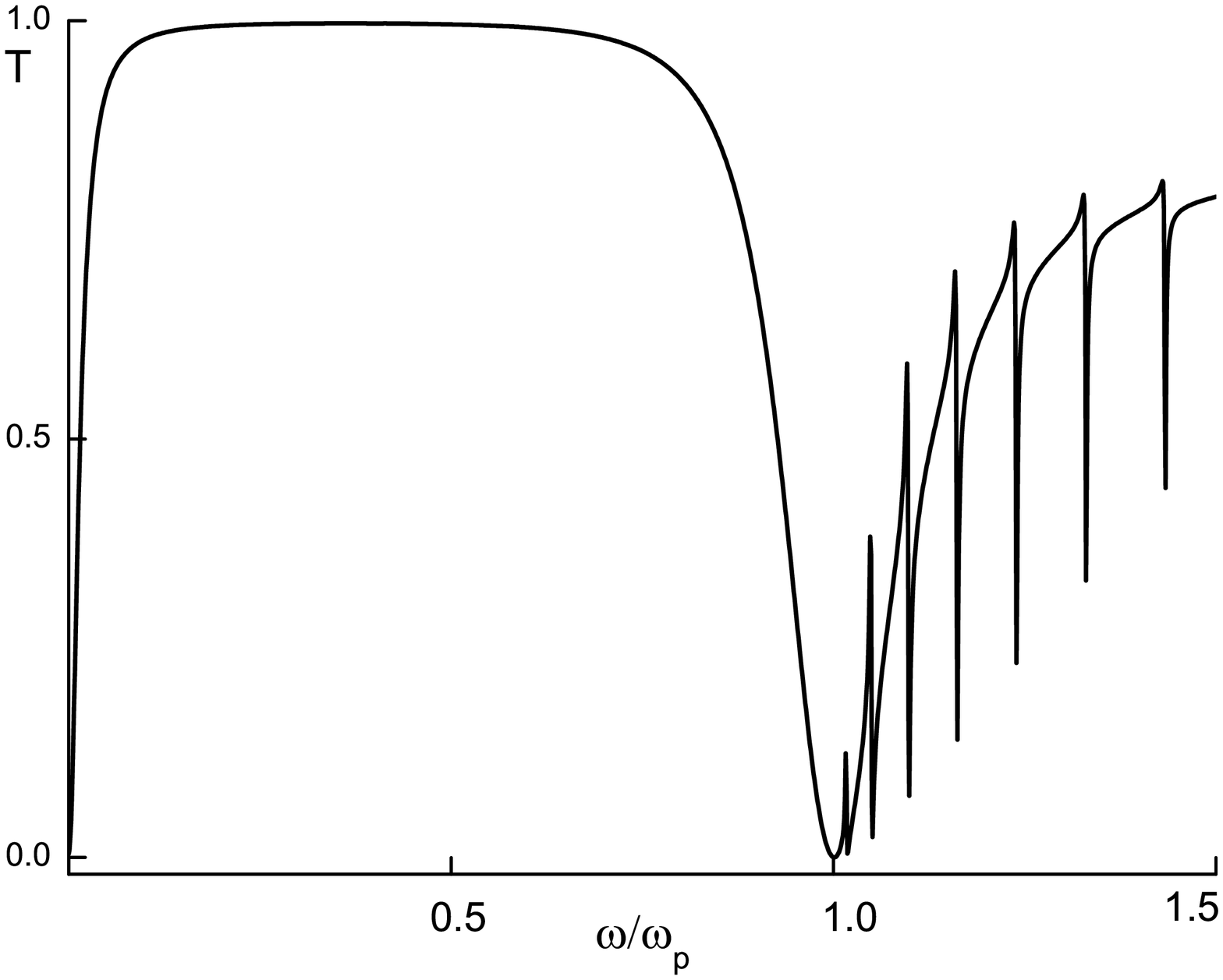}
\noindent\caption{Transmittance, $d=5$ nm,
$\nu=0.001\omega_p$, $\theta=75^\circ$.}
\includegraphics[width=16.0cm, height=6cm]{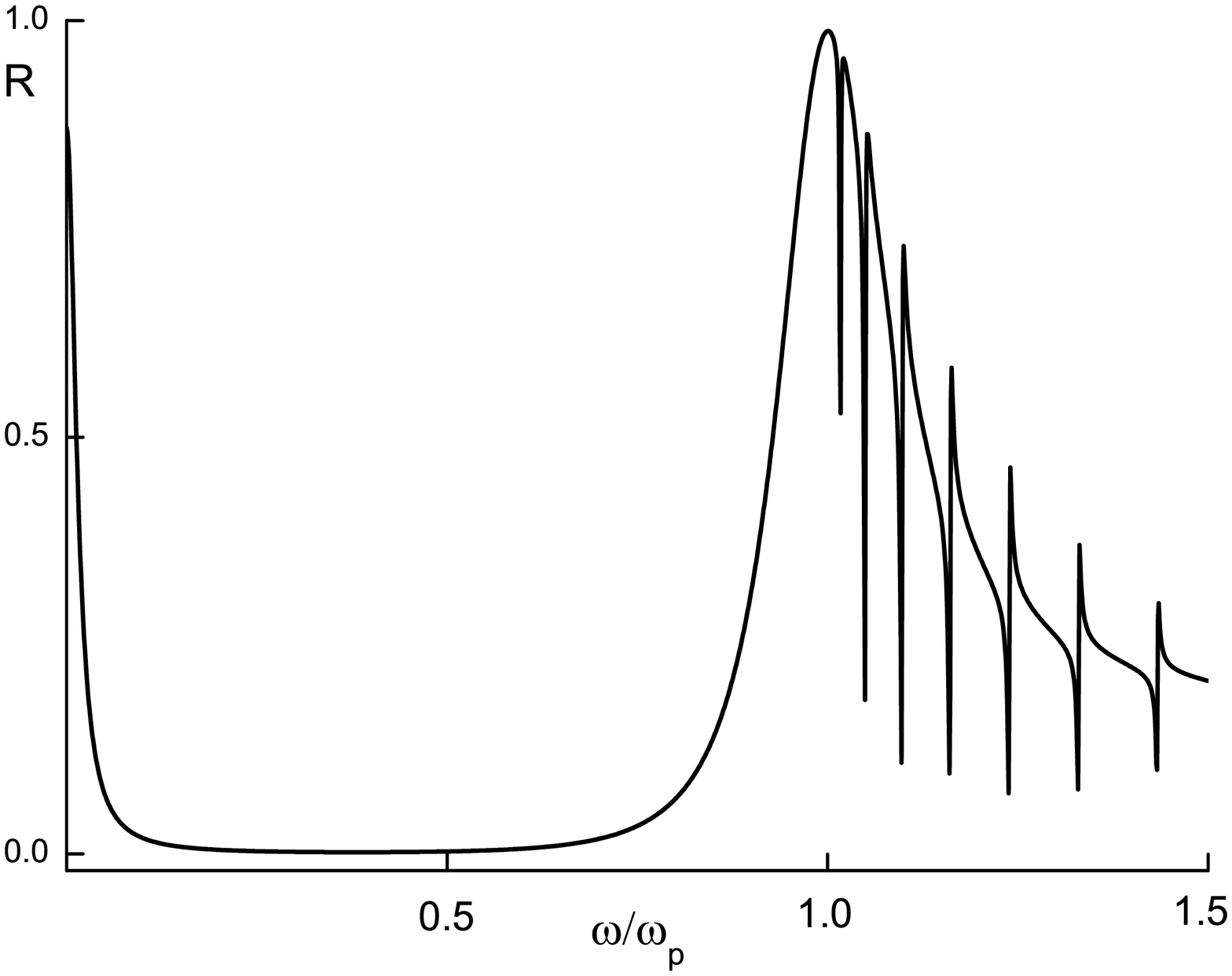}
\noindent\caption{Reflectance, $d=5$ nm,
$\nu=0.001\omega_p$, $\theta=75^\circ$.}
\includegraphics[width=16.0cm, height=7cm]{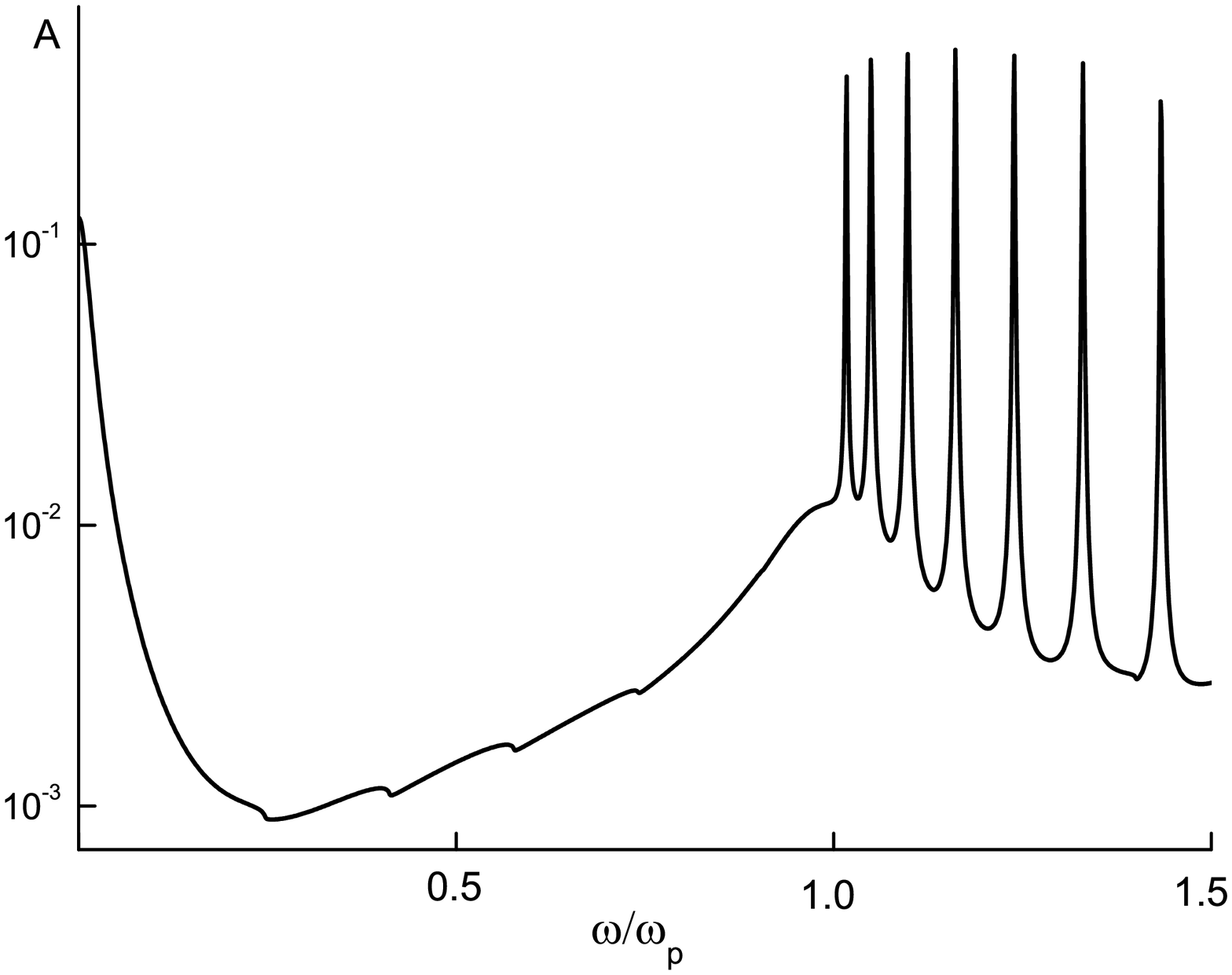}
\noindent\caption{Absorptance, $d=5$ nm, $\nu=0.001\omega_p$,
$\theta=75^\circ$.}
\end{figure}

\begin{figure}[t]\center
\includegraphics[width=16.0cm, height=6cm]{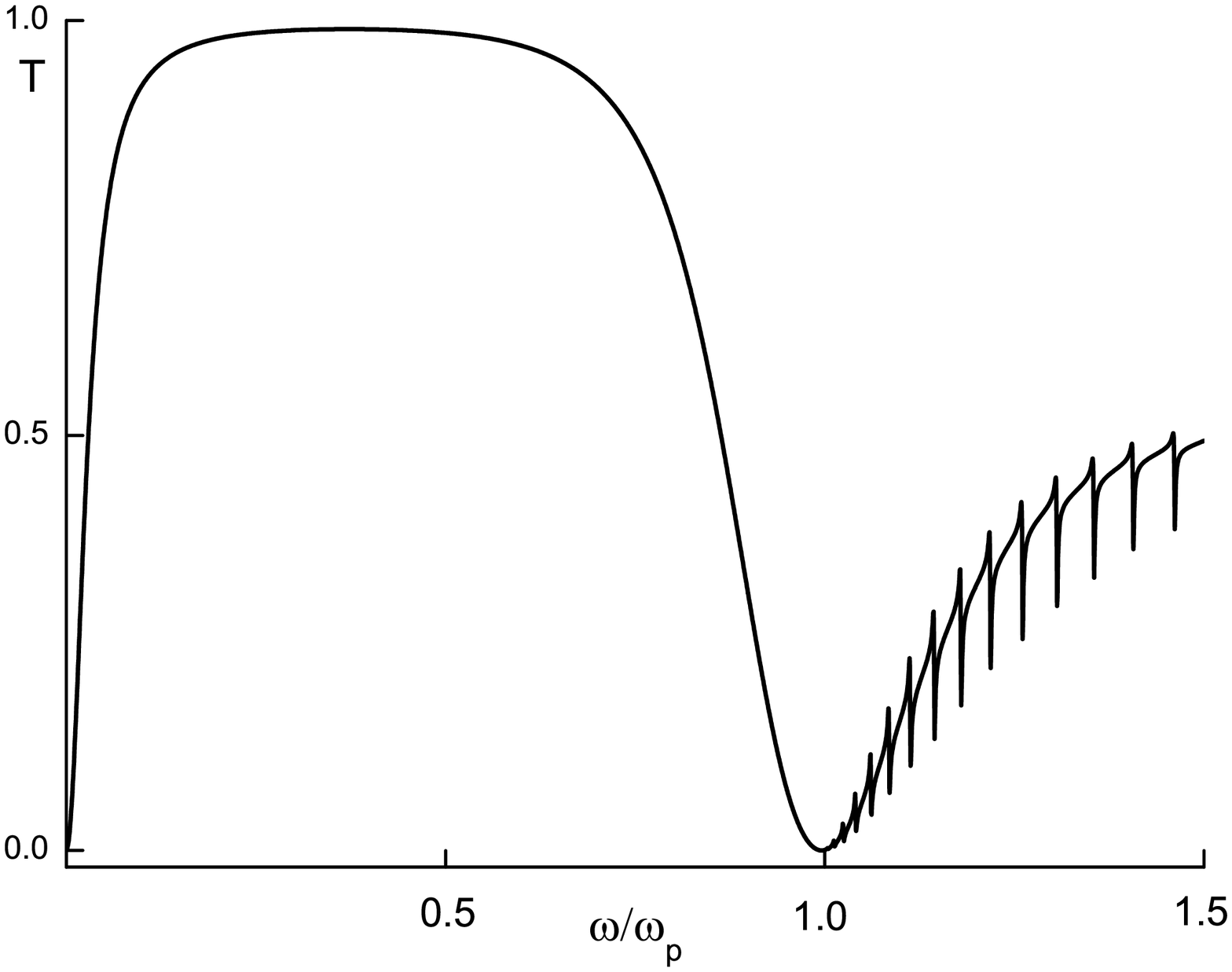}
\noindent\caption{Transmittance, $d=10$ nm,
$\nu=0.001\omega_p$, $\theta=75^\circ$.}
\includegraphics[width=16.0cm, height=6cm]{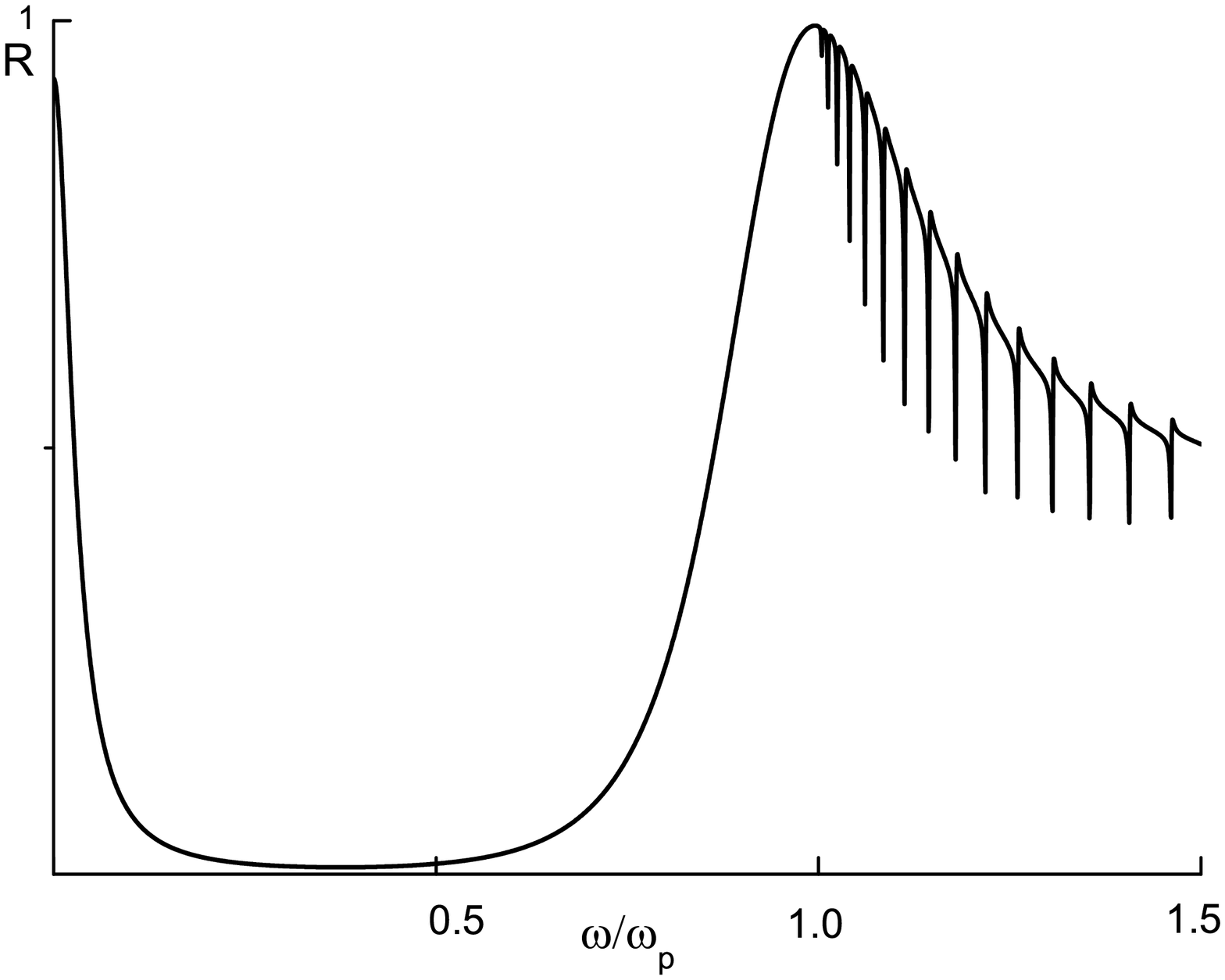}
\noindent\caption{Reflectance, $d=10$ nm,
$\nu=0.001\omega_p$, $\theta=75^\circ$.}
\includegraphics[width=17.0cm, height=7cm]{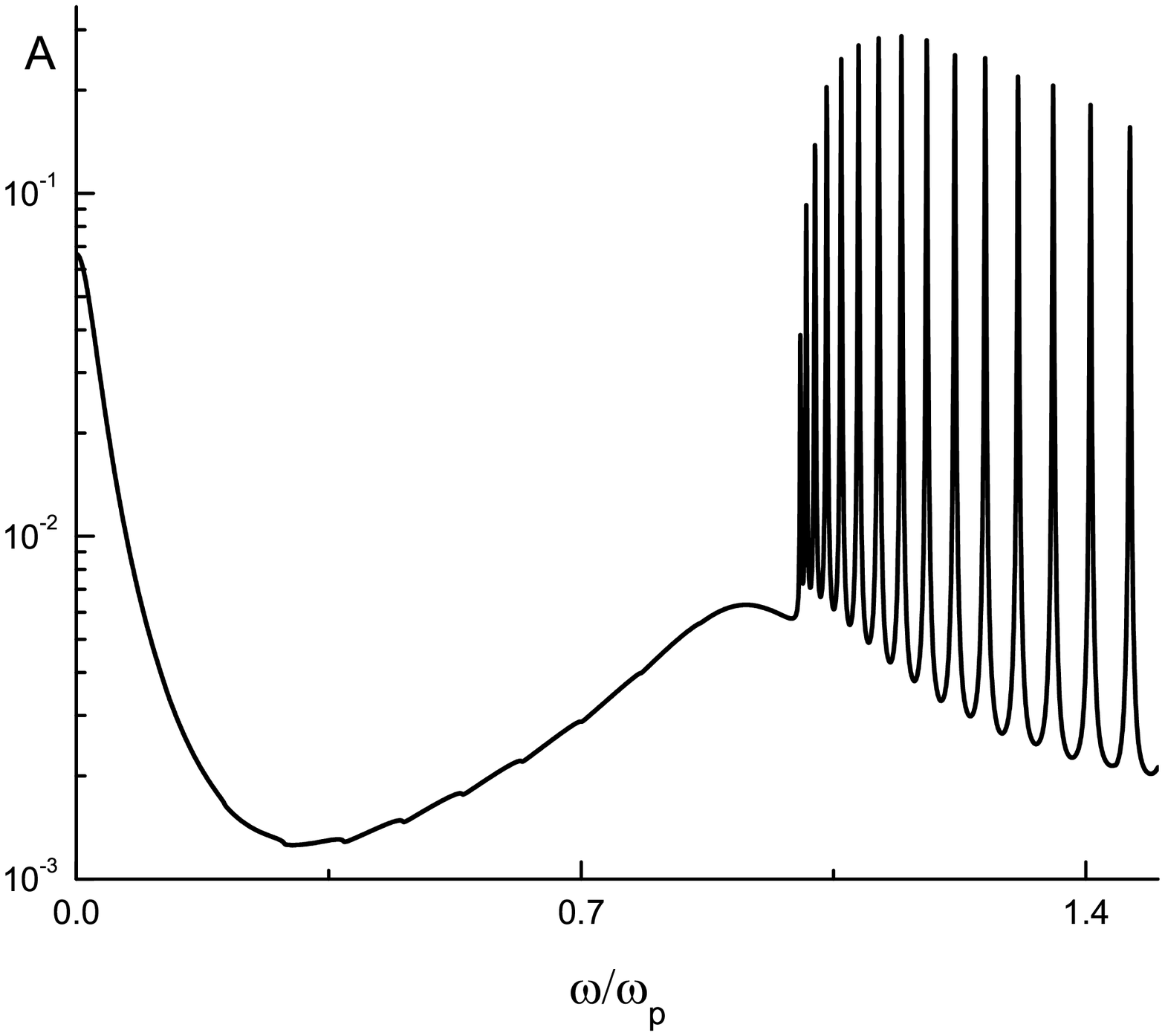}
\noindent\caption{Absorptance, $d=10$ nm,
$\nu=0.001\omega_p$, $\theta=75^\circ$.}
\end{figure}

\clearpage

At the further increase of the thickness of a film frequency of
teeths of the comb (number of links of paling) increases. Let us
note, that Fig. 10 actually coincides with Fig. 3 of work \cite
{F69}, and Fig. 9 coincides with Fig. 2 of \cite {F69}.

On Figs. 11 and 12 dependences of coefficients of transmission,
reflection and absorption from the pitch angle $\theta$  of the
electromagnetic waves on a film are presented. These graphics show
monotonously decreasing character of the transmission coefficient as
in area before resonant frequencies ($\omega<\omega_p$), and in the
field of resonant frequencies as well. The coefficient of reflecion
in the field of resonant frequencies has one minimum for thin films
(for example, for $d=1$ nm and $\nu=0.001\omega_p \;\sec^{-1}$).

On Fig. 13 we will show dependence of coefficients of transmission,
reflection and absorption on quantity of a thickness of a film $d,
\; 1$ nm $\leqslant d \leqslant 100$ nm, by $\nu=0.001 \omega_p
\;\sec^{-1}, \theta=75 ^\circ $. Transmittance has one minimum in a
considered  range of thickness, and the absorptance  has one
maximum.

The formula (24) means, that periodic character of transmittance,
reflec\-tan\-ce and absorptance explaints under the presence
function $\tanh(z_0/\eta_0)$ in the second member of this formula.
This member is called Debaye mode.

On Figs. 14 and 15 is shown, that transmittance, reflec\-tan\-ce and
absorptance have extrema in the same points $\Omega_n$,
independently of quantity of the pitch angle of the electromagnetic
wave. These reasons allow to find a thickness of a film on those
points $\Omega_n=\omega_n/\omega_p$, in which coefficients $T, R $
and $A $ have an extremum.

Let us pass to a deducing of the formula for calculation of the
thickness of a film in those points $\Omega_n $, in which
coefficients of transmission, reflection and absorption have
extrema. We consider coefficient of reflection. For this coefficient
on Fig. 16 the first links of the comb represented earlier on Figs.
8 - 10 are considered. In this figure the dot curve corresponding to
discrete and continuous spectrum, coincides with continuous curve,
answering to discrete spectrum.

Points $\Omega_n$ in which the reflection coefficient has minimum,
in accuracy coincide with points, in which function $\cos(\Re
iz_0/\eta_0)$ possesses the value of zero (see Fig. 16). From the
equation $\cos(\Re iz_0/\eta_0)=0$ we find:
$$
\Re\Big(i\dfrac{z_0(\Omega_n,\varepsilon,d)}
{\eta_0(\Omega_n,\varepsilon)}\Big)=\dfrac{\pi}{2}+\pi n,\qquad
n=0,1,2,3,\cdots,
$$
or, in explicit form,
$$
\Re\Bigg(i\dfrac{\omega_p \cdot 10^{-7}\cdot(\varepsilon-i\Omega_n)}
{2v_F\eta_0(\Omega_n,\varepsilon)}d\Bigg)=\dfrac{\pi}{2}+\pi n,\qquad
n=0,1,2,3,\cdots.
\eqno{(27)}
$$

In the formula (27) the quantity $d $ (thickness of a film) is
measured in nanometers. The exact formula for calculation of the
thickness of a film on frequencies  $\Omega_n,\; n=0,1,2, \cdots $
is deduced from the formula (27). In the points  $\Omega_n,\;
n=0,1,2, \cdots $ the reflection coefficient has local minima:
$$
d=\dfrac{10^7\pi v_F(1+2n)}
{\omega_p\Re\Big(\dfrac{\Omega_n+i\varepsilon}
{\eta_0(\Omega_n,\varepsilon)}\Big)}, \qquad n=0,1,2,3,\cdots.
\eqno{(28)}
$$

From Fig. 16 it is visible that at $n=3 \; \Omega_3=1.025$. The
formula (28) gives the thickness: $d=9.968$ nm, i.e. an error in
visual determination of the thickness of the film gives  $0.3\%$.

\begin{center}\bf
7.  Conclusion
\end{center}

In the present work for the thin films which thickness does not
surpass thickness skin layer, the formulas for calculation of
transmittance, reflectance and absorptance are received. The
analysis of these coefficients is carried out. The formula for a
finding of a thickness of a film by resonances in the field of
resonant frequencies is deduced.

\newpage

\begin{figure}[ht]\center
\includegraphics[width=14.0cm, height=6cm]{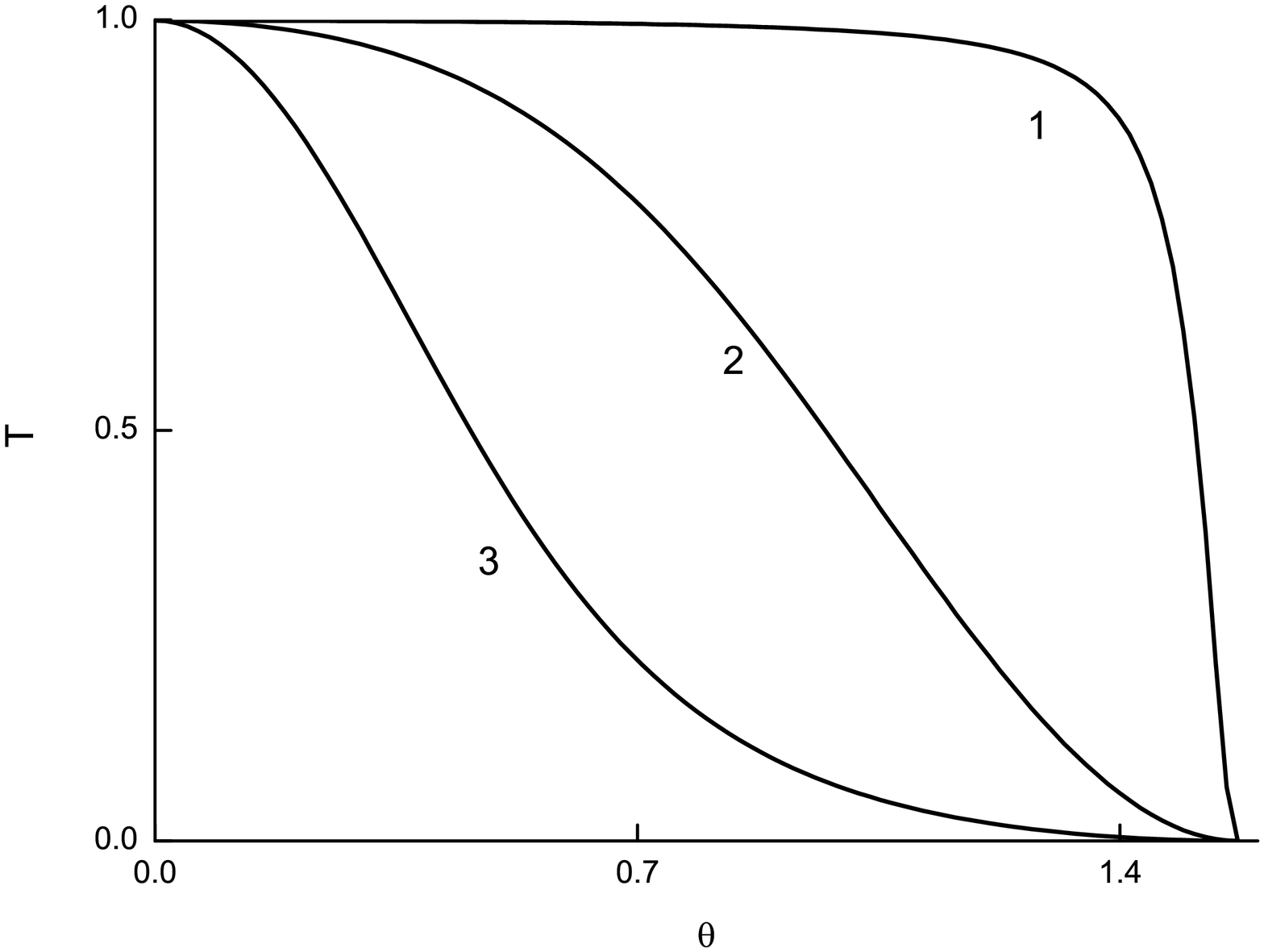}
\noindent\caption{Transmittance, $d=2$ nm,
$\nu=0.005\omega_p$. Curves $1,2,3$ correspond to
values of parameters
$\omega=0.9\omega_p,\omega=\omega_p,
\omega=1.01\omega_p$.}
\includegraphics[width=14.0cm, height=6cm]{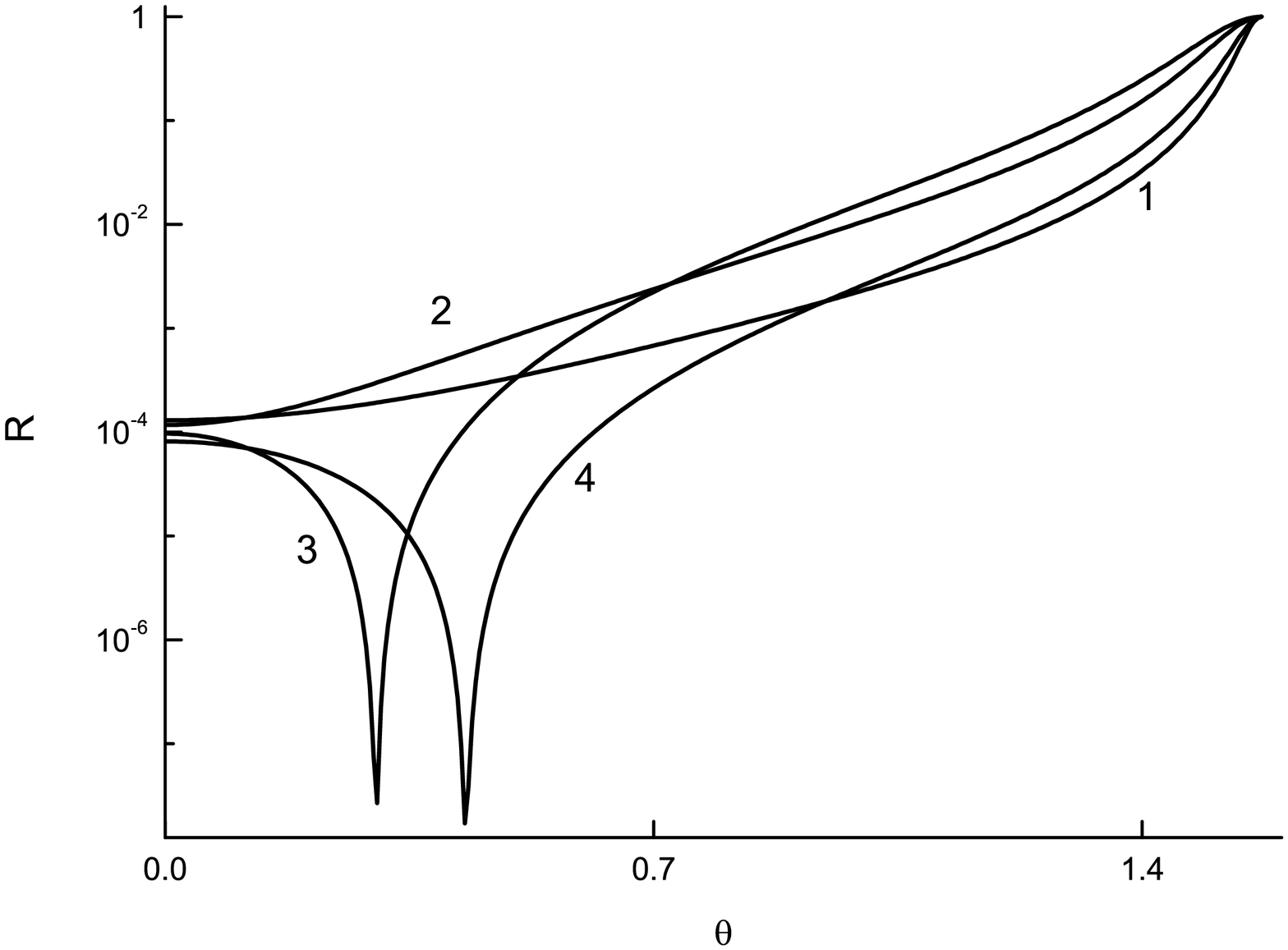}
\noindent\caption{Reflectance, $d=1$ nm,
$\nu=0.001\omega_p$. Curves $1,2,3,4$ correspond to
values of parameters
$\omega=0.95\omega_p,\omega=\omega_p,
\omega=1.1\omega_p, \omega=1.2\omega_p$.}
\includegraphics[width=14.0cm, height=7cm]{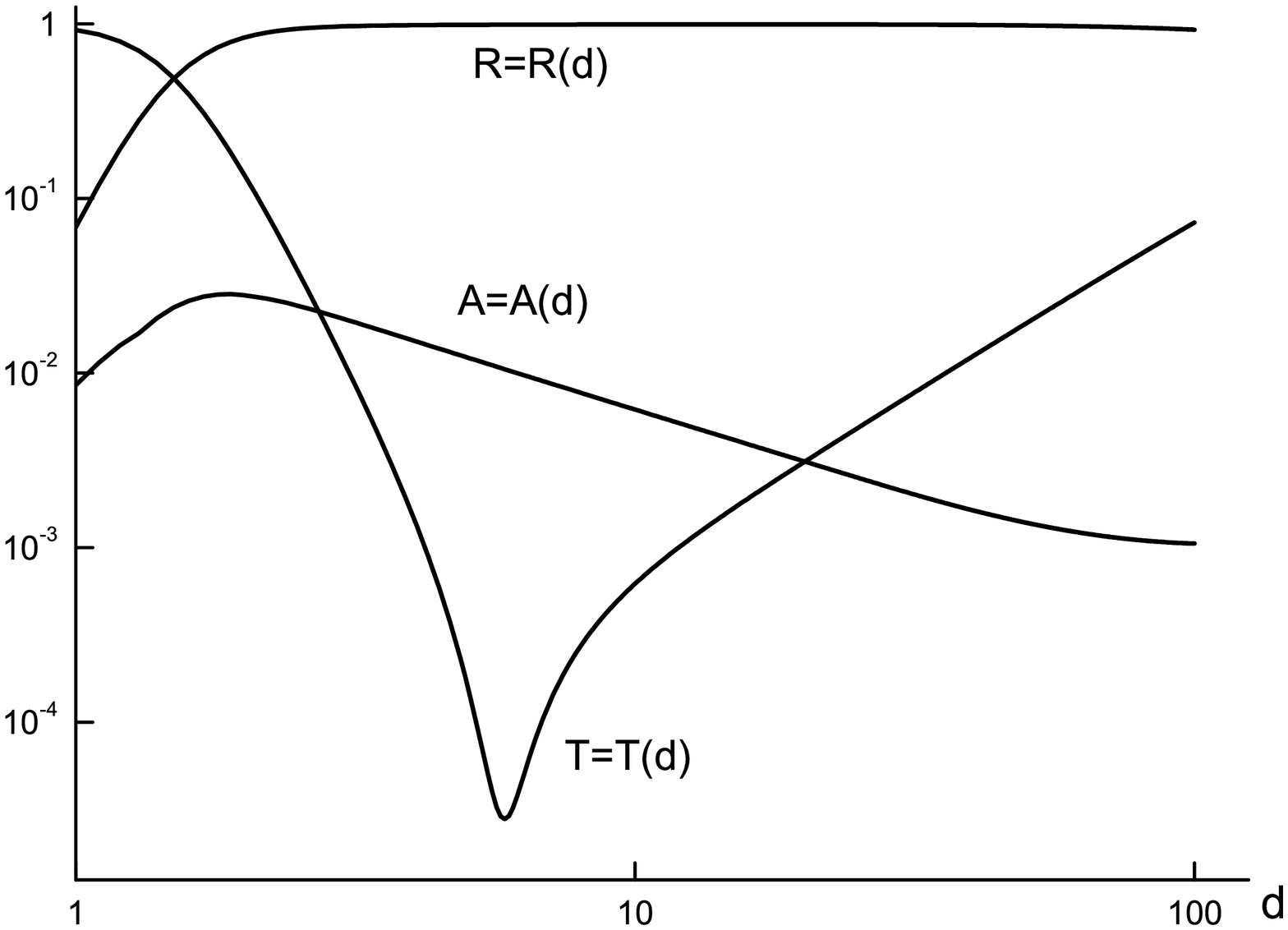}
\noindent\caption{Dependence of transmittance, reflectance and
absorptance from film thickness $d,\;1$ nm $\leqslant d \leqslant 100$ nm,
$\omega=\omega_p, \nu=0.001\omega_p$.}\label{rateIII}
\end{figure}

\begin{figure}[h]\center
\includegraphics[width=14.0cm, height=6cm]{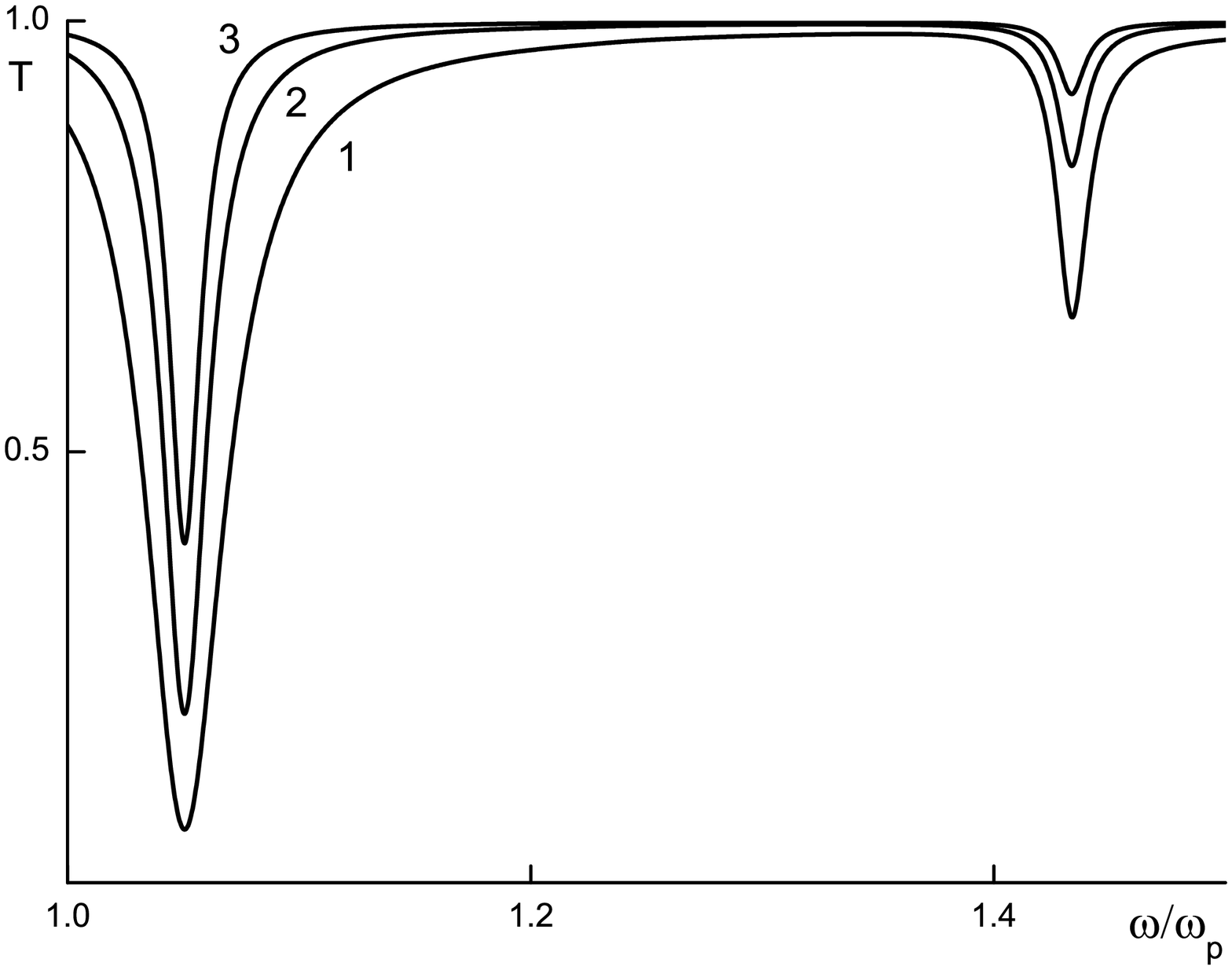}
\noindent\caption{Transmittance, $d=1$ nm,
$\nu=0.01\omega_p$. Curves $1,2,3$ correspond to
values of angle $\theta=75^\circ,60^\circ,45^\circ$.}
\includegraphics[width=14.0cm, height=6cm]{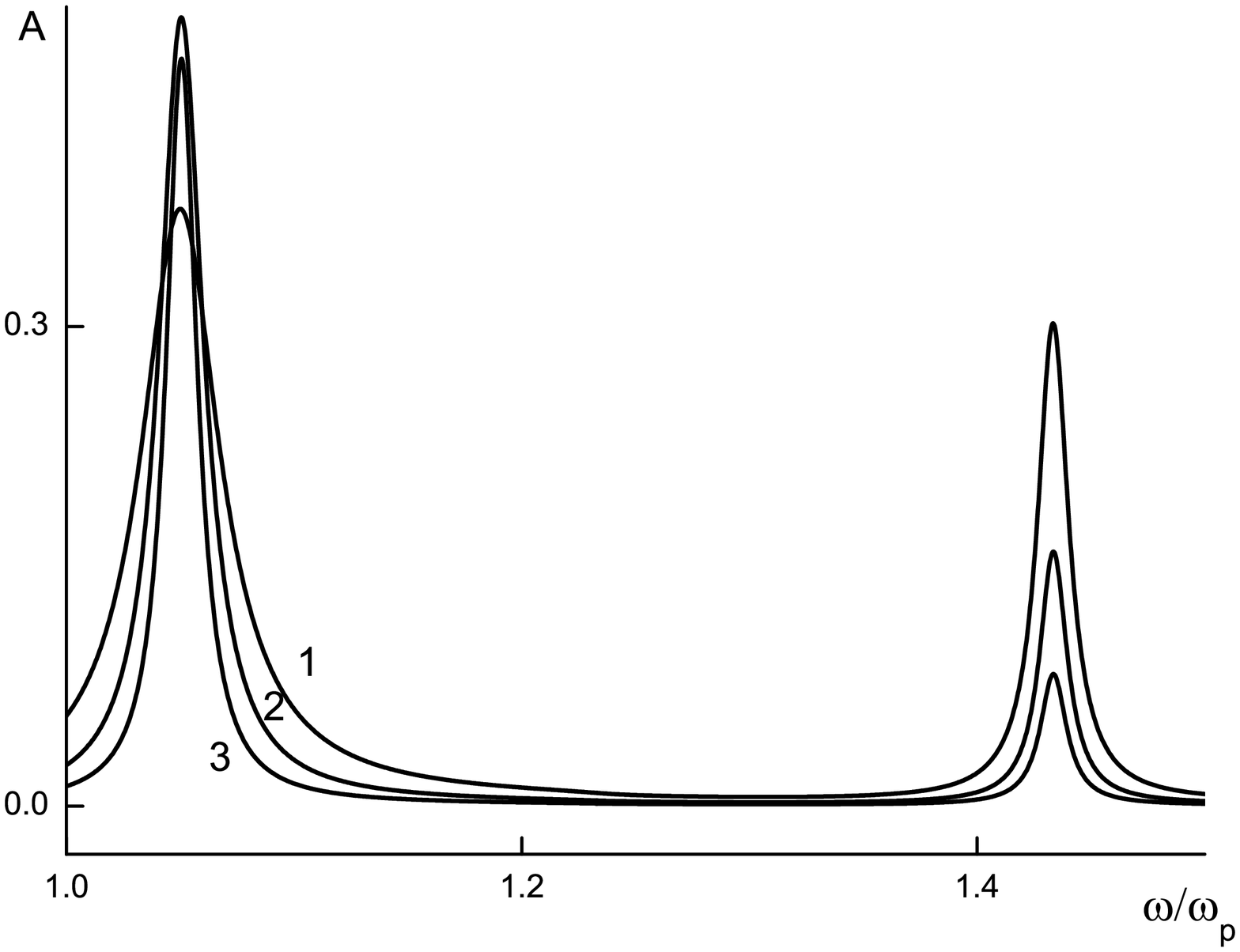}
\noindent\caption{Absorptance, $d=1$ nm,
$\nu=0.01\omega_p$. Curves $1,2,3$ correspond to
values of angle $\theta=75^\circ,60^\circ,45^\circ$.}
\includegraphics[width=18.0cm, height=7cm]{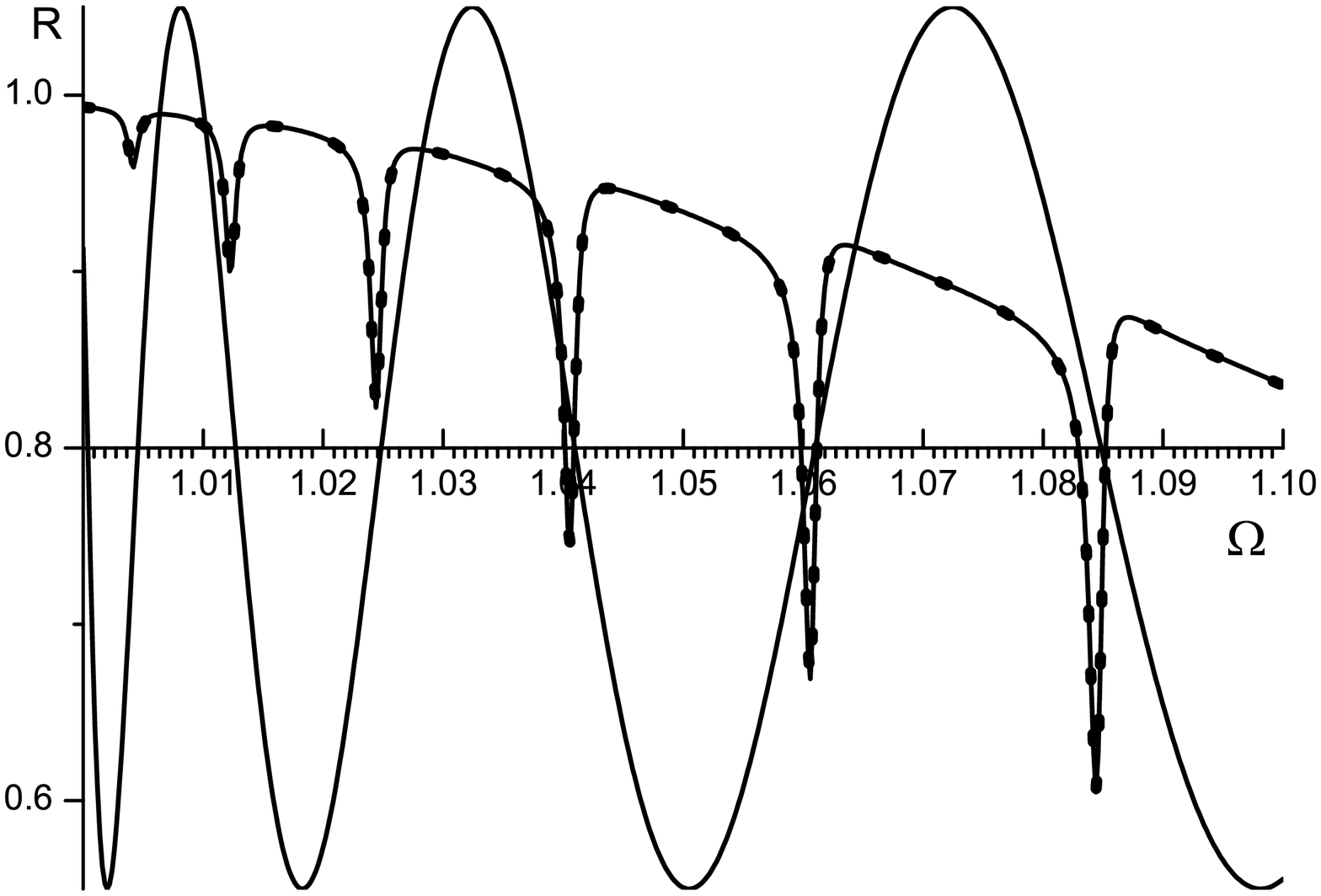}
\noindent\caption{Reflectance, $d=10$ nm,
$\nu=0.001\omega_p$. Periodic curve is set by equation
$y=0.8+0.25\cos(\Re(iz_0/\eta_0))$.}
\end{figure}

\end{document}